\documentclass{book}    
\usepackage{piers}  
\usepackage{float}

\begin{document}

\title{Topological Edge Solitons in Polaritonic Lattice}
\maketitle

\begin{authors}

{\bf D. R. Gulevich}$^{1}$, {\bf D. Yudin}$^{1}$, {\bf D. V. Skryabin}$^{1,2}$, {\bf I. V. Iorsh}$^{1}$, 
{\bf and I. A. Shelykh}$^{1,3}$\\
\medskip
$^{1}$ITMO University, St. Petersburg 197101, Russia\\
$^{2}$Department of Physics, University of Bath, Bath BA2 7AY, United Kingdom\\
$^{3}$Science Institute, University of Iceland, Dunhagi 3, IS-107, Reykjavik, Iceland

\end{authors}

\begin{paper}

\begin{piersabstract}
We study discrete nonlinear edge excitations of polaritonic kagome lattice. We show that when nontrivial topological phase of polaritons is realized, the kagome lattice permits propagation of bright solitons formed from topological edge states. 
\end{piersabstract}


\psection{Introduction}

Mapping condensed matter electronic systems to photonic structures is powerful tool for simulation of complex condensed matter phenomena since it allows to use well-controllable photonic systems to study their far less controllable solid state counterparts~\cite{Georgescu2014}. It is therefore not surprising that topological ideas have been widely explored in the domain of photonics since pioneer work of Raghu and Haldane~\cite{Raghu-2008} who brought forward the ideas of photonic chiral edge states and Berry curvature for the photonic bands. Propagation of topological edge modes was then studied in real photonic crystals~\cite{Wang-2009} and concepts of the photonic analogues of the Hall states~\cite{Wang-2009,Haldane2008a, Wang2008} and topological insulators~\cite{Hafezi-2011,Fang-2012,Rechtsman-2013,Khanikaev2013} were introduced. This allowed extensive theoretical and experimental study of the effects of non-trivial topology in electromagnetic systems in the broad spectral range from microwave to optical frequencies~\cite{Lu-2014}.

Despite this success,
optical circuits miss one important property inherent to solid state: strong inter-particle interactions. 
Exciton-polarions~\cite{Carusotto2013}, quasi-particles originating from strong coupling of quantum well-excitons and cavity photons in semiconductor microcavities, can be exploited to fill this gap. Due to their mixed light-matter nature, they can be both effectively controlled by well developed methods of optical confinement (similar to those used in photonic lattices), and demonstrate extremely strong nonlinear response (due to exciton-exciton interactions). Exciton-polaritons thus offer a unique laboratory for study of non-trivial topological phases in presence of interactions.
The emergence of the non-trivial topology and the existence of the topologically protected edge states in polaritonic lattices of different geometries has been shown theoretically in a number of works~\cite{Karzig-PRX-2015, Bardyn-PRB-2015, Nalitov-Z, Yi-PRB-2016, Bardyn-PRB-2016, Janot-PRB-2016, Gulevich-kagome}. 
Recently, the focus of attention in the study of the effects of non-trivial topology has shifted towards systems with nonlinearity~\cite{Lumer, Ostrovskaya, Hadad,  Bleu-PRB-2016, Bardyn-PRB-2016, Kart, Gulevich-Meissner, Ablowitz, Leykam, Kartashov-Optica, Soln-PRL-2017,  Gorlach-PRA-2017}, where exciton-polaritons, due to their unique properties, play a special role.
Among these works are studies of self-localized states~\cite{Lumer, Ostrovskaya}, self-induced topological transitions~\cite{Hadad}, topological Bogoliubov excitations~\cite{Bardyn-PRB-2016}, suppression of topological phases~\cite{Bleu-PRB-2016}, vortices in lattices~\cite{Kart}, spin-Meissner states in ring resonators~\cite{Gulevich-Meissner}, solitons in lattices~\cite{Ablowitz, Leykam, Kartashov-Optica, CerdaMendez2016}, dimer chains~\cite{Soln-PRL-2017}, dark solitons in kagome lattice~\cite{DGulevich-SciRep}. In the present work we present our results for propagation of solitons formed from topological edge states of polaritonic kagome lattice.


\psection{Results}

The tight-binding Hamiltonian for polaritons confined inside an array of coupled microcavity pillars forming kagome lattice reads~\cite{Gulevich-kagome}:
\begin{multline}
\hat{H}=
\Omega \sum_{i,\sigma}\sigma\,\hat{a}_{i,\sigma}^\dagger \hat{a}_{i,\sigma}
- \sum_{\langle ij\rangle,\sigma} \left( J \hat{a}_{i,\sigma}^\dagger \hat{a}_{j,\sigma} 
+ \delta J e^{2i\varphi_{ij}\bar\sigma} \hat{a}_{i,\sigma}^\dagger \hat{a}_{j,\bar\sigma} 
 + h.c.\right) \\
+ \sum_{i,\sigma}\left( 
\frac{\alpha_1}{2}\, \hat{a}_{i,\sigma}^\dagger \hat{a}_{i,\sigma}
\hat{a}_{i,\sigma}^\dagger \hat{a}_{i,\sigma} 
+ \alpha_2\, \hat{a}_{i,\sigma}^\dagger \hat{a}_{i,\sigma}
\hat{a}_{i,\bar\sigma}^\dagger \hat{a}_{i,\bar\sigma} \right).
\label{tb}
\end{multline}
Here, the operators $\hat{a}_{i,\sigma}^\dagger$ ($\hat{a}_{i,\sigma}$) create (annihilate) exciton-polariton of circular polarization $\sigma=\pm$ at site $i$ of the lattice, the summation $\langle ij\rangle$ goes over nearest neighbors (NN), the angles $\varphi_{ij}$ specify the directions of vectors connecting the neighboring sites. The first term in~\eqref{tb} describes Zeeman energy splitting ($2\Omega$) of the circular polarized components induced by the external magnetic field, the second term describes the NN hopping with conservation and inversion of circular polarization (the latter comes from TE-TM splitting), and the last term describes the on-site polariton-polariton interactions with effective constants $\alpha_1$ and $\alpha_2$. In what follows, we will use normalized units where $J$ is a unit of the energy and the interpillar distance is unit of the length. 


First, consider the linear regime when interactions in~\eqref{tb} can be neglected.
In Ref.~\cite{Gulevich-kagome} it was shown that the presence of the finite TE-TM splitting $\delta J$ and magnetic field $\Omega$ opens a gap which separates the Bloch bands into two bundles with nontrivial topology, see Fig.~\ref{fig:strip-disp}. According to the bulk-boundary correspondence there must exist topological edge states propagating along the interface of a finite (or, semi-finite) lattice. The peculiar shape of the dispersion in Fig.~\ref{fig:strip-disp} makes favourable the existence of nonlinear excitations of the edge states in the form of solitons.

\setlength{\unitlength}{0.1in}
\begin{figure*}
	\begin{center}
$
\begin{array}{cc}
		\begin{picture}(22,17)
		\put(12,16){(a)}        
		\put(0,1.5){\includegraphics[width=2.3in]{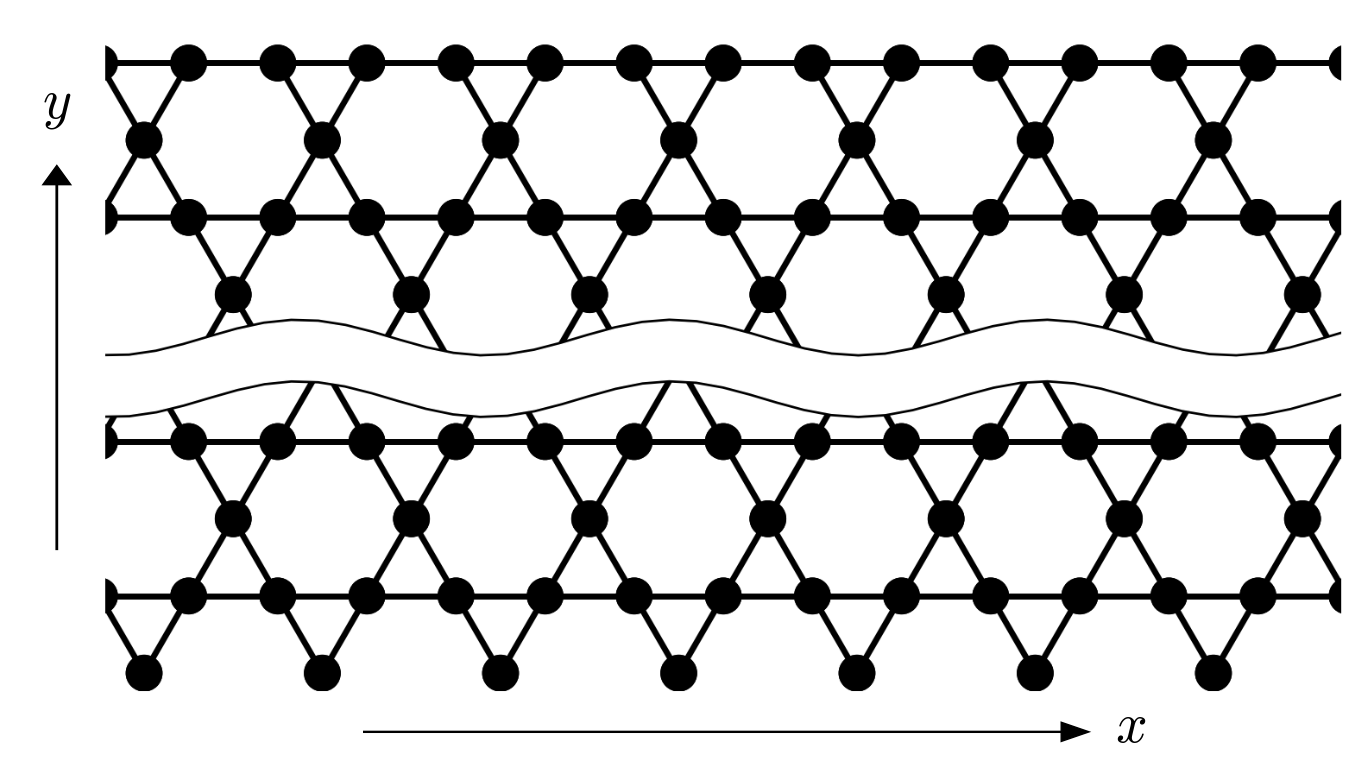}}
		\end{picture}
		&
		\begin{picture}(40,17)
		\put(19,16){(b)}                
		\put(0,0){\includegraphics[width=4.0in]{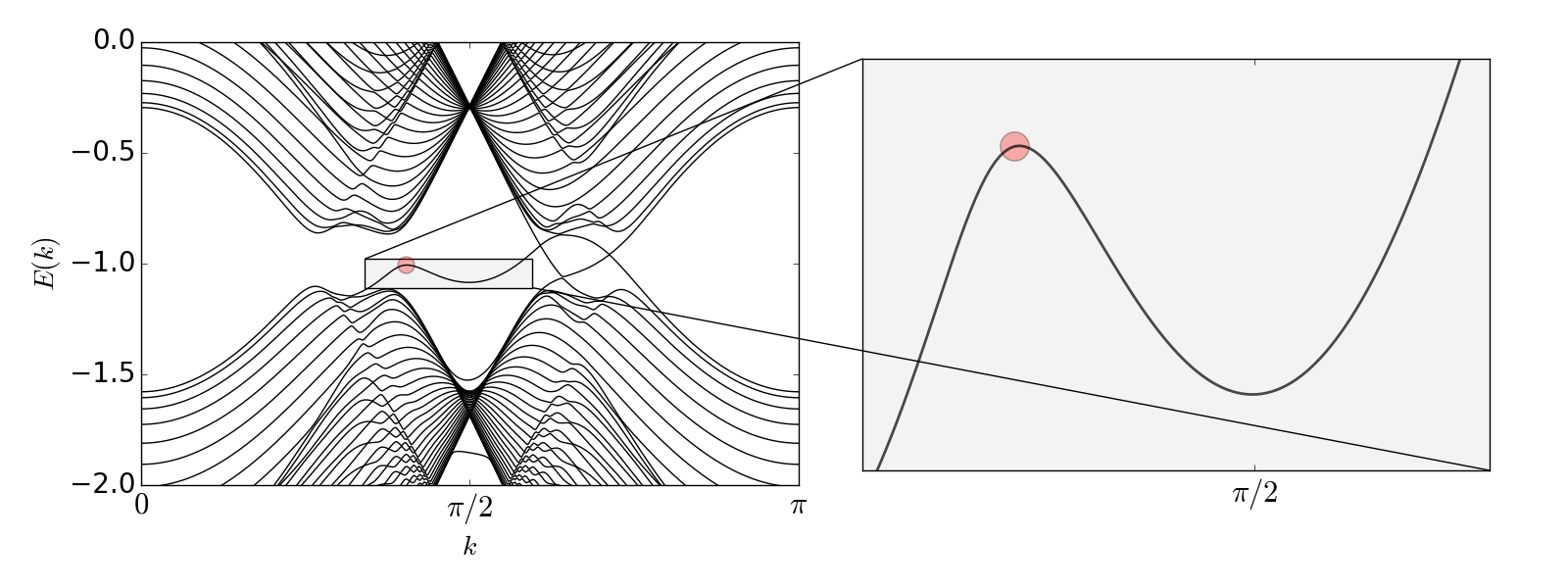}}
		\end{picture}
\end{array}
$
\caption{\label{fig:strip-disp} 
(a) The strip of polaritonic kagome lattice. The strip is infinite along $x$ and is of finite extent along $y$ direction. 
(b) Dispersion of the topological edge modes propagating in the strip of kagome lattice at $\Omega=0.3$ and $\delta J=0.15$. 
The inset on the right shows zoom of the dispersion curve where solution in the form of  bright edge soliton appears.
}
\end{center}
\end{figure*}

Numerically calculated dynamics of bright soliton formed by topological edge state 
in presence of the interactions is shown in Fig.~\ref{fig:bright}.
The initial condition for the numerical simulations was choosen as bell-shape profile with fixed quasimomentum in the vicinity of the maximum of the dispersion in Fig.1. For the specifically chosen shape of the initial excitation and nonlinearity strength, the excitation keeps its form and starts propagating along the boundary. The speed and the direction of the propagation depend on the chosen value of quasimomentum of the topological edge dispersion.

\setlength{\unitlength}{0.1in}
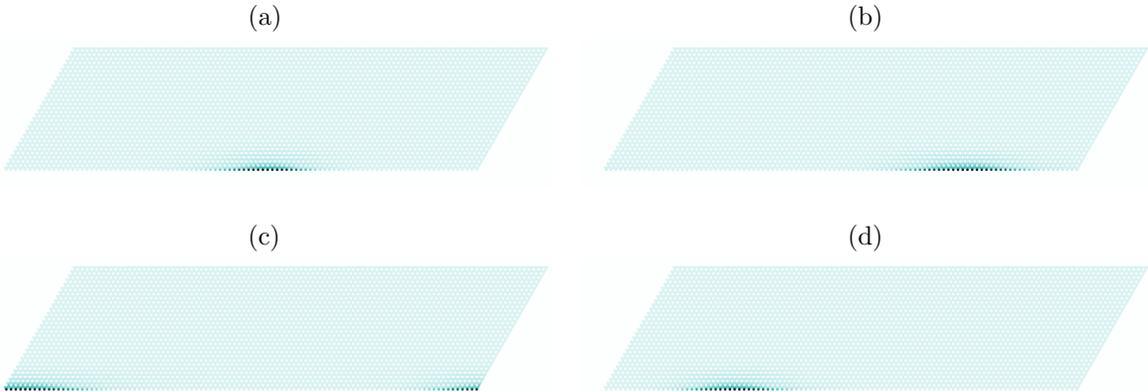
\begin{figure*}
	\begin{center}
$
\begin{array}{cc}
		\begin{picture}(30,11)
		\put(0,0){\includegraphics[width=3.0in]{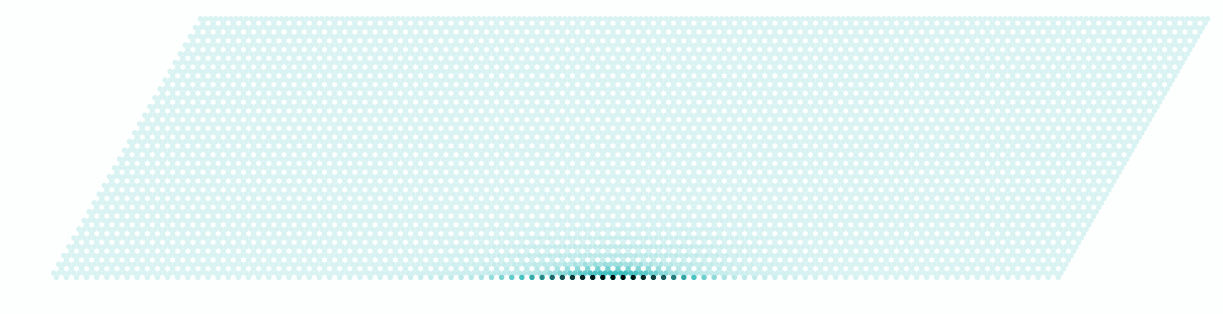}}
		\put(14,8.5){(a)}
		\end{picture}
		&
		\begin{picture}(30,11)
		\put(0,0){\includegraphics[width=3.0in]{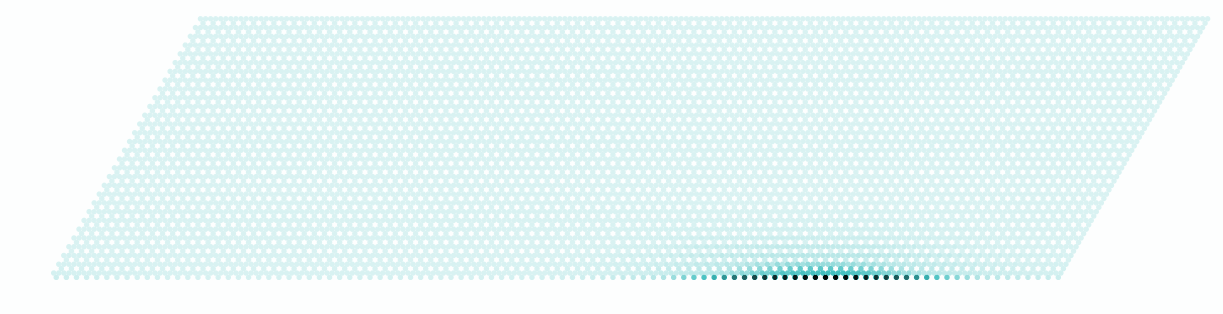}}
		\put(14,8.5){(b)}
		\end{picture}
		\\
		\begin{picture}(30,11)
		\put(0,0){\includegraphics[width=3.0in]{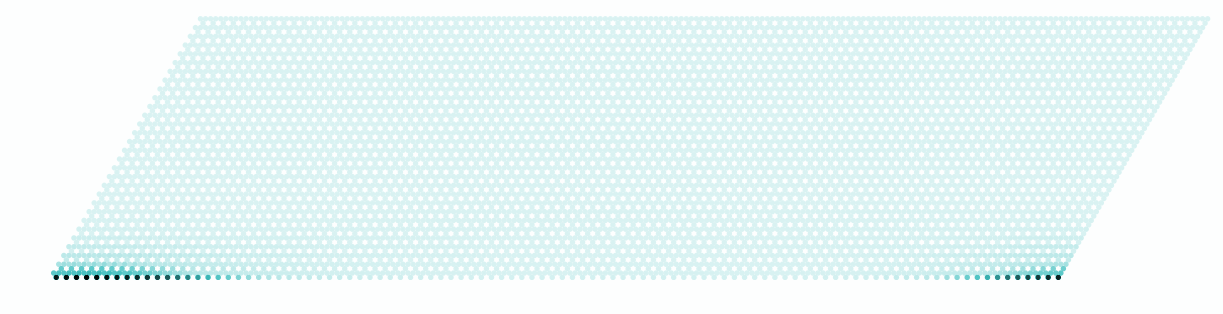}}
		\put(14,8.5){(c)}
		\end{picture}
		&
		\begin{picture}(30,11)
		\put(0,0){\includegraphics[width=3.0in]{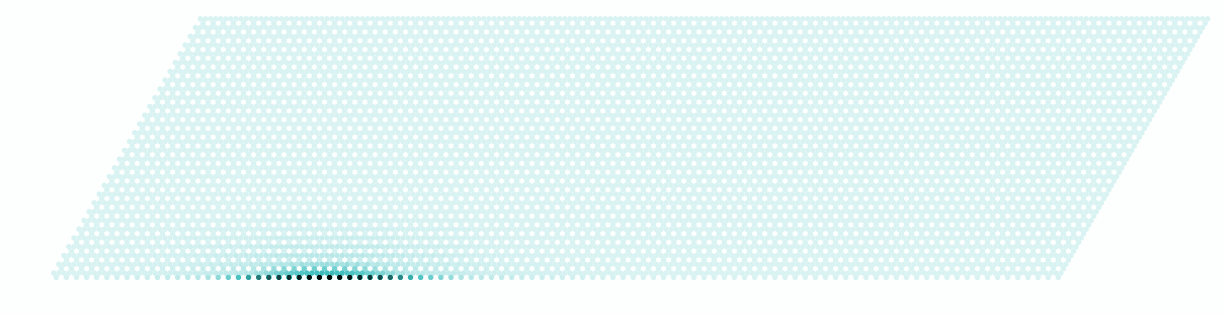}}
		\put(14,8.5){(d)}
		\end{picture}		
\end{array}
$
\caption{\label{fig:bright} 
Propagation of a bright soliton in a strip of kagome lattice. Figures (a,b,c,d) represent snapshots of the numerically calculated dynamics at subsequent time intervals. Periodic boundary conditions are imposed along the horizontal axis.
}
\end{center}
\end{figure*}

\psection{Conclusion}
We demonstrated propagation of bright solitons formed from topological edge states. Such solitons are formed from wavepackets with quasimomentum in the vicinity of the local maximum of the edge state dispersion.

\psection{Acknowledgements}

D.R.G., D.Y. and I.V.I. acknowledge support from the grant 3.8884.2017   of   the   Ministry   of   Education   and   Science of the Russian Federation.
D.Y. acknowledges support from RFBR project 16-32-60040. I.A.S. acknowledge support from the Icelandic Research Fund, Grant No. 163082-051, mega-grant № 14.Y26.31.0015 of the Ministry of Education and Science of the Russian Federation and Horizon2020 RISE project CoExAN, Grant No. 644076. 


\end{paper}

\end{document}